\documentclass[prd,preprint,tightenlines,superscriptaddress,endfloats*]{revtex4b5}

\usepackage{bm}
\usepackage{xspace}
\usepackage{graphicx}

\newcommand{\GeV}{{\ensuremath{\rm GeV}}\xspace}
\newcommand{\GeVc}{\ensuremath{{\rm GeV}/c}\xspace}
\newcommand{\MeV}{\ensuremath{{\rm MeV}}\xspace}
\newcommand{\fb}{\ensuremath{{\rm fb}}\xspace}

\newcommand{\X}{\enspace}

\begin{document}


\title{
	\begin{flushleft}\includegraphics[width=3cm]{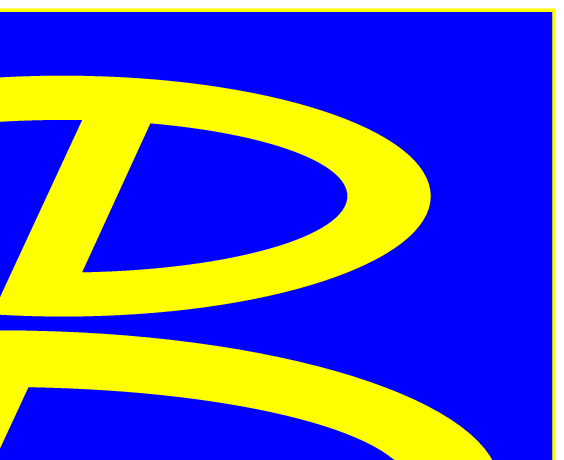}\end{flushleft}
	\vspace{-3cm}
	\vbox{\normalsize%
        \noindent%
        \rightline{\hfill {\tt KEK preprint 2001-2}}\\%
        \rightline{\hfill {\tt Belle preprint 2001-3}}%
        } 
	\vspace{3cm}
	Measurement of Inclusive Production of Neutral Pions from
        $\bm{\Upsilon(4S)}$ Decays
\begin{center}\vspace{10mm}{\normalsize Belle Collaboration}\end{center}}

%
%
%
\author{K.~Abe}
\affiliation{High Energy Accelerator Research Organization (KEK), Tsukuba}
\author{K.~Abe}
\affiliation{Tohoku University, Sendai}
\author{I.~Adachi}
\affiliation{High Energy Accelerator Research Organization (KEK), Tsukuba}
\author{Byoung~Sup~Ahn}
\affiliation{Korea University, Seoul}
\author{H.~Aihara}
\affiliation{University of Tokyo, Tokyo}
\author{M.~Akatsu}
\affiliation{Nagoya University, Nagoya}
\author{G.~Alimonti}
\affiliation{University of Hawaii, Honolulu HI}
\author{K.~Aoki}
\affiliation{High Energy Accelerator Research Organization (KEK), Tsukuba}
\author{K.~Asai}
\affiliation{Nara Women's University, Nara}
\author{Y.~Asano}
\affiliation{University of Tsukuba, Tsukuba}
% ...

\collaboration{Belle Collaboration}
\noaffiliation 

\date{March 24, 2001}

\begin{abstract}
Using the Belle detector operating at the KEKB $e^+e^-$ storage ring,
we have measured the mean multiplicity
and the momentum spectrum of neutral pions from the decays
of the $\Upsilon(4S)$ resonance.  We measure a mean of
$4.70 \pm 0.04 \pm 0.22$ neutral pions per $\Upsilon(4S)$ decay.
%
\end{abstract}


\maketitle


This analysis presents the first results
on the average multiplicity of neutral pions and their momentum spectrum in
$B$ meson decays at the $\Upsilon(4S)$ resonance using the
Belle detector~\cite{ref-belle-detector} at KEKB~\cite{ref-KEKB}.
The measurement is based on the sample of $3.4 \times 10^6$
$B$ meson pairs collected by the Belle detector at
a center-of-mass (CM) energy of $\sqrt{s}=10.58\,\GeV$.

  The particle composition of hadronic final states in $e^+e^-$ annihilation
and the measurement of inclusive particle production rates have been
important subjects for various energy regions.
Previous studies of particle composition
at the $\Upsilon(4S)$ resonance include measurements of charged particles
($\pi^\pm$~\cite{ref-CLEO-pipm,ref-ARGUS-pipm} and
$K^\pm$~\cite{ref-CLEO-kpm,ref-ARGUS-kpm}),
$\eta$ mesons~\cite{ref-CLEO-eta},
and 
vector mesons~\cite{ref-CLEO-phi,ref-ARGUS-vect}.
For typical $B$ meson decays, the bulk
of the neutral energy is carried by neutral pions.
%
Measurements of inclusive spectra contain information on
$B$ meson decay mechanisms, especially in the high momentum region
where important rare decays may become detectable.
If the details of these inclusive spectra are known more precisely,
they will allow better estimation of backgrounds and better modeling of
$B$ meson decay.

\begin{figure}
  \centering
  \includegraphics[bb=94 250 744 750, width=8.6cm]{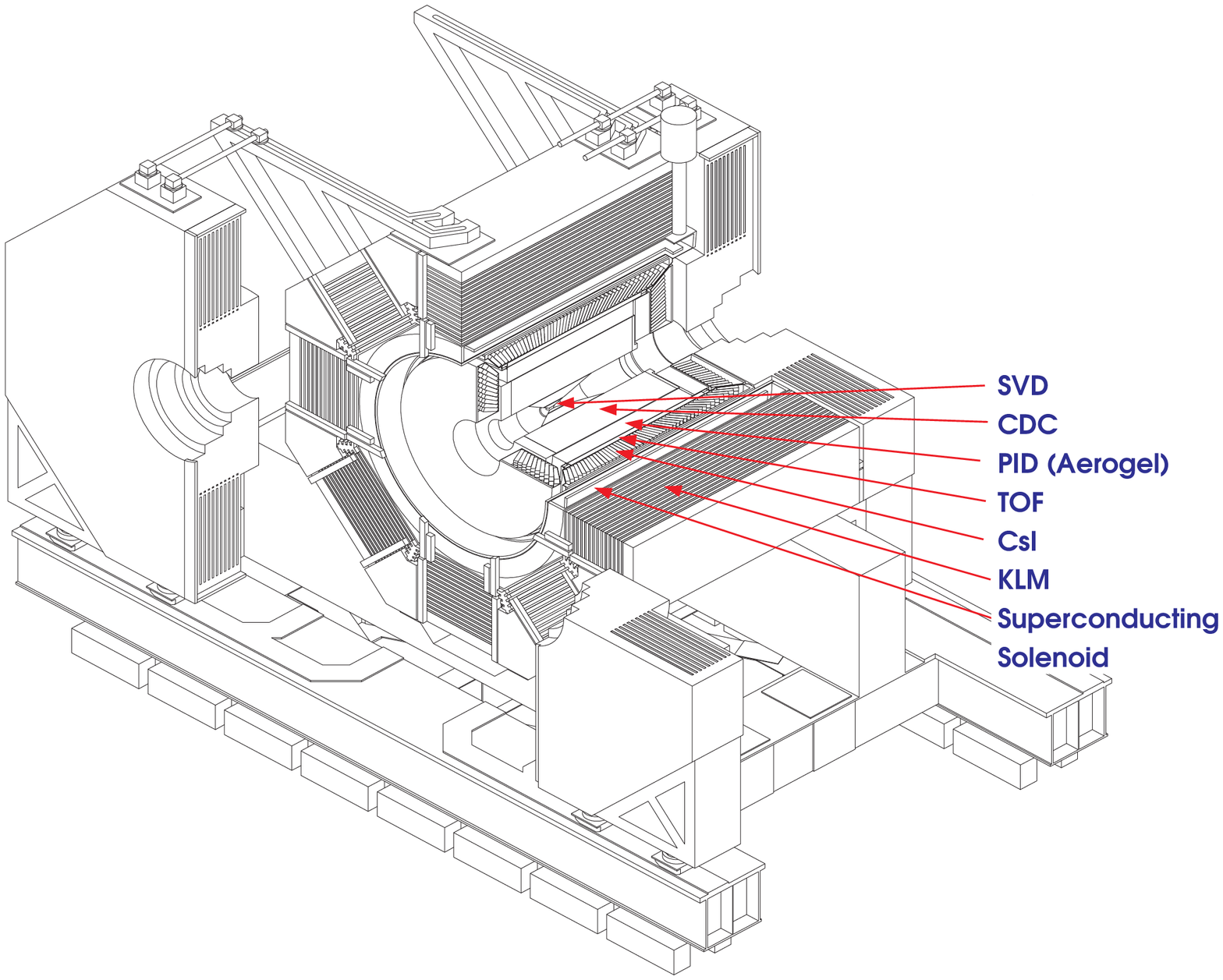}
  \caption{Isometric cutaway view of the Belle detector.
	}
  \label{fig:belle-detector}
\end{figure}

The Belle detector (see Fig.~\ref{fig:belle-detector}),
described in detail in Ref.~\cite{ref-belle-detector},
consists of a silicon vertex detector (SVD)~\cite{ref-belle-svd},
central drift chamber (CDC)~\cite{ref-belle-cdc},
aerogel \v{C}erenkov counter (ACC)~\cite{ref-belle-acc},
time of flight/trigger scintillation counter (TOF/TSC)~\cite{ref-belle-tof},
CsI electromagnetic calorimeter (ECL)~\cite{ref-belle-ecl}
and K$_L$/muon detector (KLM)~\cite{ref-belle-klm}.
The SVD measures the precise position of decay vertices.
It consists of three layers of double-sided silicon strip detectors (DSSD)
in a barrel-only design and covers 86$\%$ of the solid angle.
The layer radii are 3.0, 4.5, and 6.0 cm.
Charged tracks are reconstructed primarily by the CDC
that covers the 17$^\circ<\theta<$150$^\circ$ polar angular region.
It consists of 50 cylindrical layers of drift cells organized
into 11 super-layers (axial or small-angle-stereo), each containing
between three and six layers. A low $Z$ gas mixture (50\% He, 50\% C$_2$H$_6$)
is used to
minimize multiple-Coulomb scattering. The inner and outer radii
of the CDC are 9 cm and 86 cm, respectively. The solenoidal magnetic field
of 1.5 Tesla is chosen to optimize momentum resolution without
sacrificing reconstruction efficiency for low momentum tracks. 
Kaon identification (KID) is provided by specific ionization
($dE$/$dx$) measurements in the CDC, \v{C}erenkov threshold measurements
in the ACC, and the cylindrical TOF scintillator barrel.
The ECL is made of finely segmented CsI(T$\ell$) crystals
30 cm in length. The cross section of one counter is
approximately $55 \times 55$ mm$^2$ at the front surface.
The ECL crystals cover the 12$^\circ<\theta<157^\circ$ angular region. 
The inner radius of the barrel is 1.25 m, while the annular endcaps are
placed at +2.0 m and $-1.0$ m along the beam line from the
interaction point.  The calibration of the calorimeter is
performed using cosmic rays and Bhabha events.
The KLM consists of
alternating layers of glass resistive plate counters and 4.7 cm thick iron plates.

The data samples used in this analysis correspond to
$3.2\,\fb^{-1}$
of integrated luminosity taken at the $\Upsilon(4S)$
resonance and $0.6\,\fb^{-1}$
taken at a CM energy 60 MeV below the resonance;
the latter was used to subtract underlying continuum background.
The integrated luminosity was determined from the number of Bhabha events
for which we require both electron and positron in the region
of $46.7^\circ<\theta^\ast<145.7^\circ$ in the center-of-mass frame.


Hadronic events are selected based on charged track information from the
CDC and cluster information from the ECL.  We require at least three
charged tracks, that the energy sum in the calorimeter be between 10\% and
80\% of $\sqrt{s}$, and that the charged track momentum be balanced in the $z$
direction.  This removes the majority of two photon, radiative Bhabha, and
$\tau^+\tau^-$ events where both $\tau$'s decay to leptons. Radiative
Bhabha events with one electron outside of the ECL acceptance are removed by
requiring at least one large-angle cluster in the ECL and requiring that the
average cluster energy be below 1 GeV.
Higher multiplicity $\tau^+\tau^-$ events are removed
if the charged and neutral energy sums in the event are consistent with
a $\tau$ pair event and if the reconstructed invariant mass of
the particles found in each of the two hemispheres perpendicular
to the event thrust axis falls below the $\tau$ mass.
Beam-gas and beam-wall backgrounds are removed by reconstructing the primary
vertex of the event and requiring it to be consistent with the known
location of the interaction point.
The selection is 99\% efficient for $B\bar{B}$ events
and approximately 87\% efficient for continuum events.
The efficiency for the hadronic event selection
was found by a GEANT-based~\cite{Brun:1987ma} Monte Carlo simulation program.
To suppress continuum, we require that the ratio $R_2$ of
second to zeroth Fox-Wolfram moments~\cite{ref-fwm},
determined using charged tracks and neutral clusters,
be less than 0.5.


Photons are reconstructed from neutral clusters
in the ECL that have a lateral shape consistent with that
of an electromagnetic shower.
The energy resolution was measured to be
$\sigma_E/E = 0.066\%/E \oplus 0.81\%/E^{0.25} \oplus 1.34\%$ ($E$ in GeV)
from beam tests~\cite{ref-belle-ecl}.
To keep the combinatorial background at a reasonable level,
only photons in the central barrel region
(35$^{\circ}<\theta<$120$^{\circ}$) with
$E_{\gamma} \ge 30\,\MeV$ are used in this analysis; the endcap regions have
worse energy resolution due to more intervening 
material and higher beam-associated background.
\begin{figure}
\centering
\includegraphics[width=8.6cm]{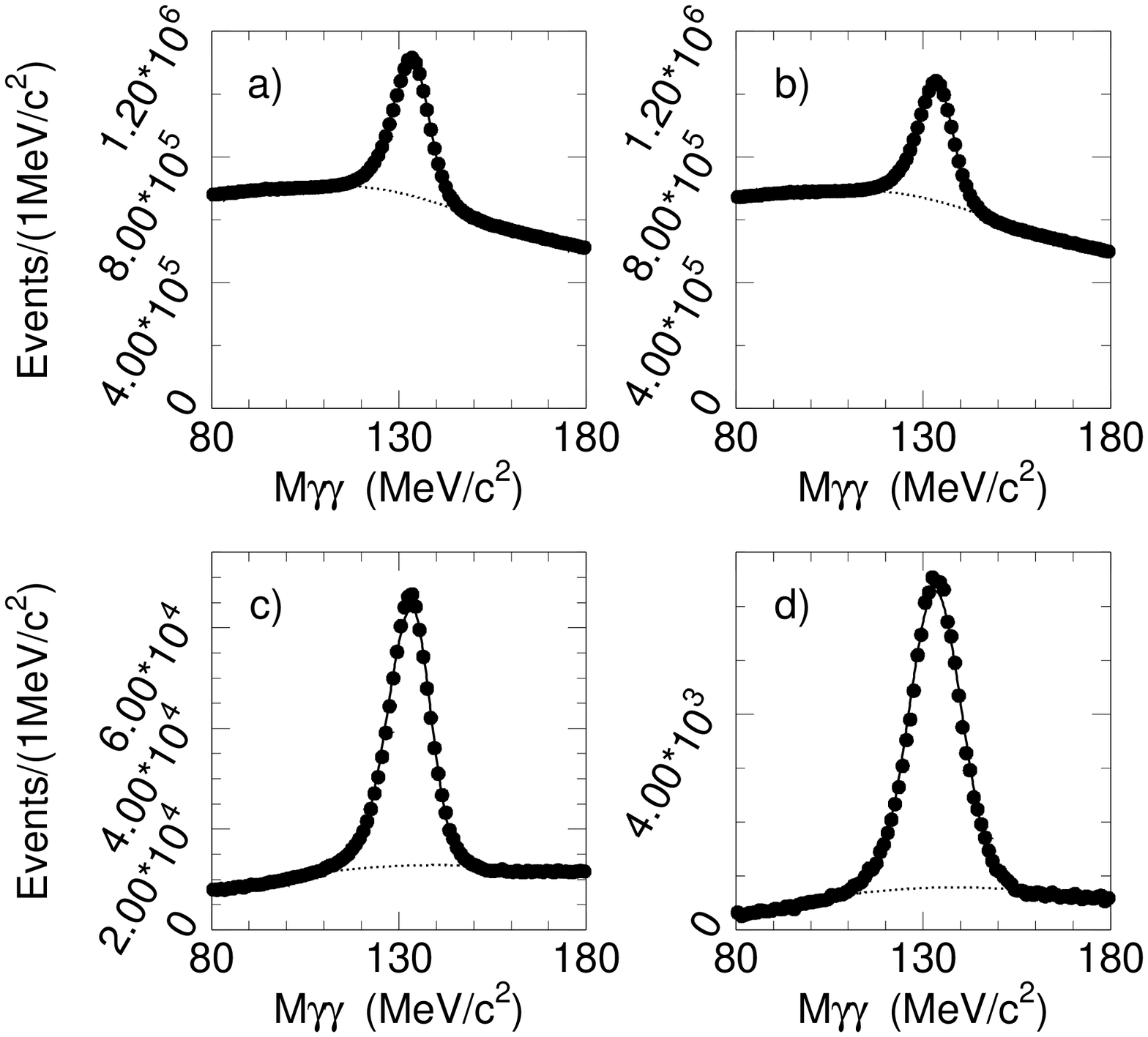}
\caption{$\gamma\gamma$ invariant mass distributions for the $\Upsilon(4S)$ resonance data. (a) $p^{\ast}_{\gamma\gamma} = $ 0.0--3.0 GeV/$c$; (b) 0.0--1.0 GeV/$c$; (c) 1.0--2.0 GeV/$c$; (d) 2.0--3.0 GeV/$c$. An average mass resolution of 5 MeV/$c^2$ was obtained. The smooth curve in each plot is a fit to the data using an asymmetric (or symmetric) Gaussian plus a polynomial.}
\label{fig:mgg}
\end{figure}

\begin{figure}
  \centering
  \includegraphics[width=8.6cm]{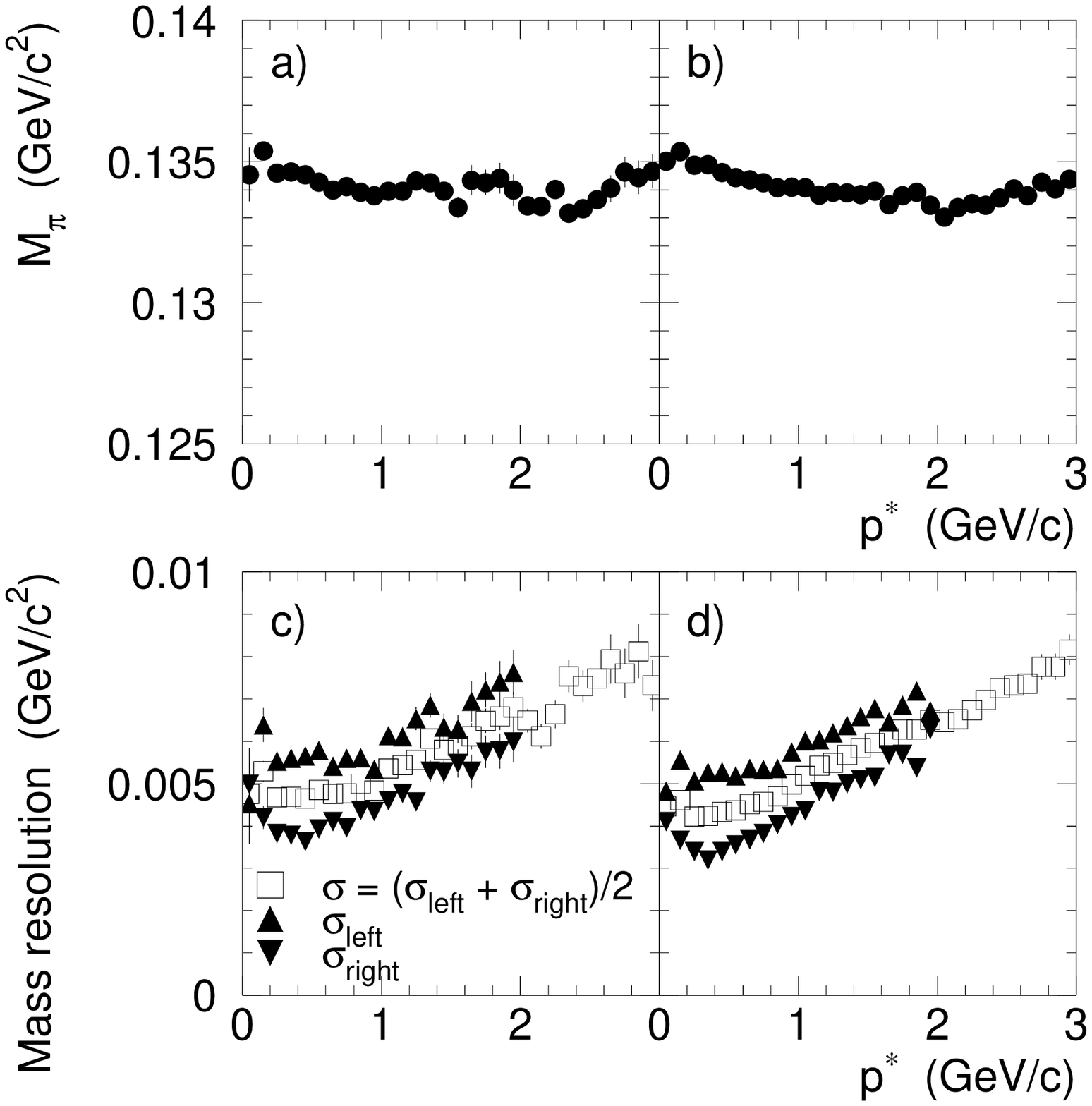}
  \caption{Mass peak position and resolution obtained from off-resonance data
	and Monte Carlo expectations.
	(a),(c) data; (b),(d) continuum Monte Carlo.
	The fluctuations near 2 GeV/$c$ are caused by the change in the
	signal fitting function from an asymmetric to a symmetric Gaussian.
	}
  \label{fig:mass-data-mc}
\end{figure}

For each 100 MeV/$c$ momentum bin in the CM momentum range 0 to 3 GeV/$c$,
the $\gamma\gamma$ invariant mass distribution is fit to an 
asymmetric (symmetric above 2 GeV/$c$) Gaussian (i.e., a Gaussian
with different widths on either side of the mean)
for the signal plus a polynomial for the combinatorial background to extract
the $\pi^{0}$ yields. 
Fig.~\ref{fig:mgg} shows typical mass spectra obtained from
the on-resonance data.
For the asymmetric Gaussians, the mass resolution is defined
as the mean of the left- and right-hand sigmas.
An average mass resolution of 5 
MeV/$c^2$ is obtained,
dominated by energy resolution at low momenta or by angular
resolution at high momenta.  The mass peak is shifted slightly from
the 
established
value~\cite{ref-pdg} because of the
asymmetric energy response of the calorimeter due to
shower leakage. The observed mass peak position and resolution are
consistent with Monte Carlo expectations, as shown
in Fig.~\ref{fig:mass-data-mc}.

To extract the $\pi^0$ momentum
spectrum from $\Upsilon(4S)$ decays, the underlying
continuum in the on-resonance data
is subtracted bin-by-bin using off-resonance data.
The inclusive spectrum is calculated using
\begin{equation}
  \frac{1}{\sigma_{h}} \cdot \frac{d\sigma_{\pi^{0}}}{dp^{\ast}_{\pi^{0}}}
  = \frac{1}{N_{h}}
    \cdot 
    \frac{Y^{i}_{\rm on} - \alpha \cdot Y^{i}_{\rm off}}
         {\epsilon^{i} \cdot \Delta p^{\ast}_{\pi^{0}}} \ ,
\label{eq:cross-section}
\end{equation}
where  $N_{h}$ is the number of produced $B\bar{B}$ events,
$Y_{\rm on}$ and $Y_{\rm off}$ are the 
background-subtracted $\pi^0$
yields obtained from on- and off-resonance data fits,
$\alpha = ({\cal L}_{\rm on}/{\cal L}_{\rm off}) \cdot (s_{\rm off}/s_{\rm on})$
is the on-off scaling factor,
and $\epsilon$ is the product of acceptance and detection efficiency
for each momentum bin.
The average $\pi^0$ multiplicity is obtained by summing the data
from the measured individual momentum bins as shown in
Table~\ref{tab-spectrum}.

The $\pi^0$ acceptance and detection efficiency are determined from
Monte Carlo simulations of $B\bar{B}$ decays
(equal proportions of charged and neutral $B$ mesons)
and continuum processes.
The $\gamma\gamma$ invariant mass distributions are fit with the same
functions as used in the real data analysis.  The product of acceptance
and detection efficiency are defined as the ratio of the fitted $\pi^0$
yield to the generated count.
Efficiencies for the high momentum $\pi^0$'s above the kinematic limit
for $B\bar{B}$ decays are deduced from continuum
Monte Carlo normalized in the 2.2--2.3 GeV/$c$ bin, and scaled efficiencies
are used that account for the different acceptances.

\begingroup
\squeezetable
\begin{table}
\caption{Measured inclusive $\pi^0$ spectrum from $\Upsilon(4S)$ decays,
using $3.2\,\fb^{-1}$ on-resonance and $0.6\,\fb^{-1}$ off-resonance
data.}
\label{tab-spectrum}
\begin{ruledtabular}
\begin{tabular}{l rrr r r}
$p^{\ast}$ (GeV/$c$)&
$Y_{\rm on} - \alpha Y_{\rm off}$&
 $\epsilon$ (\%)&
 $\frac{\displaystyle 1}{\displaystyle \sigma_{h}}
\frac{\displaystyle d\sigma}{\displaystyle dp^{\ast}} \pm \Delta_{stat} \pm \Delta_{syst}$
\\\hline
0.0-0.1	&	289901.7 $\pm$ \X9256.5&	18.6&	4.483 $\pm$ 0.143 $\pm$ 0.644
\\
0.1-0.2	&	492207.2 $\pm$ 16451.8&	15.9&	8.893 $\pm$ 0.297 $\pm$ 1.182
\\
0.2-0.3	&	465198.9 $\pm$ 10435.8&	16.5&	8.106 $\pm$ 0.182 $\pm$ 1.012
\\
0.3-0.4	&	461727.6 $\pm$ \X8139.7&	20.0&	6.635 $\pm$ 0.117 $\pm$ 0.836
\\
0.4-0.5	&	411074.4 $\pm$ \X6470.8&	22.5&	5.245 $\pm$ 0.083 $\pm$ 0.699
\\
0.5-0.6	&	323401.5 $\pm$ \X5305.1&	24.8&	3.749 $\pm$ 0.062 $\pm$ 0.506
\\
0.6-0.7	&	255770.0 $\pm$ \X4409.2&	26.4&	2.780 $\pm$ 0.048 $\pm$ 0.377
\\
0.7-0.8	&	187946.7 $\pm$ \X3761.8&	27.8&	1.942 $\pm$ 0.039 $\pm$ 0.290
\\
0.8-0.9	&	139720.8 $\pm$ \X3179.7&	29.3&	1.370 $\pm$ 0.031 $\pm$ 0.209
\\
0.9-1.0	&	106438.9 $\pm$ \X2754.1&	30.1&	1.017 $\pm$ 0.026 $\pm$ 0.161
\\
1.0-1.1	&	\X79213.3 $\pm$ \X2004.2&	32.4&	0.701 $\pm$ 0.018 $\pm$ 0.125
\\
1.1-1.2	&	\X61164.2 $\pm$ \X1744.7&	32.8&	0.535 $\pm$ 0.015 $\pm$ 0.095
\\
1.2-1.3	&	\X46737.0 $\pm$ \X1528.1&	33.3&	0.402 $\pm$ 0.013 $\pm$ 0.076
\\
1.3-1.4	&	\X34688.4 $\pm$ \X1358.3&	33.9&	0.294 $\pm$ 0.012 $\pm$ 0.059
\\
1.4-1.5	&	\X26208.0 $\pm$ \X1236.0&	33.7&	0.223 $\pm$ 0.011 $\pm$ 0.049
\\
1.5-1.6	&	\X18556.8 $\pm$ \X1115.8&	35.0&	0.152 $\pm$ 0.009 $\pm$ 0.039
\\
1.6-1.7	&	\X15916.3 $\pm$ \X\X994.5&	35.5&	0.129 $\pm$ 0.008 $\pm$ 0.031
\\
1.7-1.8	&	\X12728.9 $\pm$ \X\X915.9&	36.1&	0.101 $\pm$ 0.007 $\pm$ 0.026
\\
1.8-1.9	&	\X10759.3 $\pm$ \X\X805.8&	37.0&	0.084 $\pm$ 0.006 $\pm$ 0.022
\\
1.9-2.0	&	\X\X7842.7 $\pm$ \X\X746.8&	36.7&	0.061 $\pm$ 0.006 $\pm$ 0.018
\\
2.0-2.1	&	\X\X7241.3 $\pm$ \X\X667.3&	35.2&	0.059 $\pm$ 0.005 $\pm$ 0.017
\\
2.1-2.2	&	\X\X4945.7 $\pm$ \X\X586.1&	32.4&	0.044 $\pm$ 0.005 $\pm$ 0.014
\\
2.2-2.3	&	\X\X2956.1 $\pm$ \X\X533.1&	31.2&	0.027 $\pm$ 0.005 $\pm$ 0.011
\\
2.3-2.4	&	\X\X-345.5 $\pm$ \X\X505.8&	30.7&	-0.003 $\pm$ 0.005 $\pm$ 0.011
\\
2.4-2.5	&	\X\X\X163.3 $\pm$ \X\X443.3&	29.7&	0.002 $\pm$ 0.004 $\pm$ 0.009
\\
2.5-2.6	&	\X\X\X488.3 $\pm$ \X\X415.3&	28.4&	0.005 $\pm$ 0.004 $\pm$ 0.007
\\
2.6-2.7	&	\X\X-242.0 $\pm$ \X\X394.3&	27.2&	-0.003 $\pm$ 0.004 $\pm$ 0.007
\\
2.7-2.8	&	\X\X\X189.8 $\pm$ \X\X347.3&	25.5&	0.002 $\pm$ 0.004 $\pm$ 0.006
\\
2.8-2.9	&	\X\X-206.9 $\pm$ \X\X314.9&	23.7&	-0.003 $\pm$ 0.004 $\pm$ 0.005
\\
2.9-3.0	&	\X\X\X-59.5 $\pm$ \X\X278.8&	21.6&	-0.001 $\pm$ 0.004 $\pm$ 0.005
\end{tabular}
\end{ruledtabular}
\end{table}
\endgroup


Possible sources of systematic uncertainties and their effect on the
inclusive $\pi^0$
mean multiplicity measurement are summarized
in Table ~\ref{tab-syst}.  These are discussed next.

\begin{table}
\caption{Sources of systematic uncertainty in the inclusive $\pi^{0}$ multiplicity measurement.}
\label{tab-syst}
\begin{ruledtabular}
\begin{tabular}{lc}
Source	&
Effect on $\langle n_{\pi^0} \rangle$ (\%) \\
\hline\hline
Overall normalization&	
1.5
\\
\hline
$\Delta n = Y_{\rm on} - \left(\frac{{\cal L}_{\rm on}}{{\cal L}_{\rm off}}\right)\left(\frac{s_{\rm off}}{s_{\rm on}}\right) \cdot Y_{\rm off}$&	
1.5\\
$R_2$&
0.4\\
OFF $E^{\ast}_{\gamma} \cdot \sqrt{\frac{s_{\rm on}}{s_{\rm off}}}$&
0.3\\
\hline
$E_{\gamma}$ non-linearity&	
0.6\\
$E_{\gamma}$ smearing&
2.0\\
Shower shape	&	
0.4\\
Track match	&	
1.6\\
Fit procedure	&	
3.0\\
Energy dependence of efficiency&	
0.3\\
\hline\hline
Total (added in quadrature) &	4.6
\\
\end{tabular}
\end{ruledtabular}
\end{table}

The uncertainties in the number of $B\bar{B}$
events and the hadronic event selection efficiency are estimated to be
1\% and 1.1\%, respectively, or 1.5\% combined.

The effect of uncertainty in the relative luminosity ratio,
used to subtract the continuum background from the on-resonance data,
was studied by varying the size of the continuum subtraction
by 1\%, and we find a 1.5\% uncertainty in the mean multiplicity.

Due to the CM energy difference between on- and off-resonance data,
the characteristics of continuum events may not match,
leading to a bias in the continuum subtraction from the on-resonance data.
In particular, the event shape at lower CM energy
is slightly less 
jet-like so that more continuum events survive the common
$R_2$ cut (resulting in an oversubtraction from the on-resonance data),
and the particle momenta scale with CM energy.
Hadronic event selection and
$R_2$ cut bias from the on-off energy difference were determined
by using a continuum Monte Carlo~\cite{Sjostrand:1992mh}
sample generated at the CM energy of the off-resonance data;
we estimate a 0.4\% uncertainty in the $\pi^0$ multiplicity due to this effect.
Particle momentum scaling with CM energy was
studied by comparing the multiplicity with and without momentum
scaling; this leads to an uncertainty of 0.3\%.

The uncertainties in the $\pi^0$
detection efficiency due to the application of the shower
transverse shape cut, the charged-track veto,
photon energy smearing, and the
ECL's modest non-linear energy response correction,
are estimated to be 0.4\%, 1.6\%, 2\%, and 0.6\%,
respectively (determined by comparing the yields with and without
each of these criteria).

The decay angular distribution for the photons in the $\pi^0$
rest frame is expected to be isotropic, but the detected
distribution of the decay angle $\theta_d$---
defined as the angle between the photon momentum in the $\pi^0$
rest frame and the $\pi^0$ momentum in the laboratory frame---
shows an energy dependence due to the $\gamma$ energy cut.
Inaccuracy
in the Monte Carlo simulation could manifest itself as an anisotropy 
in the efficiency-corrected decay angle distribution.
A comparison between Monte Carlo and data for
$|\cos\theta_d| > 0.5$ and $|\cos\theta_d| \le 0.5$
shows no discrepancy and is used to estimate
a systematic uncertainty of 0.3\%.

The $\gamma\gamma$ invariant mass distribution for each CM momentum bin is
fit to the following functions in the mass window 80--180 MeV/$c^{2}$:
\begin{itemize}
  \item $p^{\ast}_{\gamma\gamma} < 1.0$ GeV/$c$: 4th order polynomial plus asymmetric Gaussian;
  \item $1.0 \le p^{\ast}_{\gamma\gamma} < 2.0$ GeV/$c$: 2nd order polynomial plus asymmetric Gaussian;
  \item $2.0 \le p^{\ast}_{\gamma\gamma} < 3.0$ GeV/$c$: 2nd order polynomial plus Gaussian.
\end{itemize}
The fitting range and the order of the polynomial are chosen to minimize
statistical fluctuation.  By changing the order of the polynomial,
the relative yield variation between data and Monte Carlo
is taken as the systematic uncertainty in the background modeling;
a 2.8\% uncertainty was deduced.  The weighted average of the
fit errors in the efficiency estimation,
$(\delta \epsilon)^2 =
         \sum_{i} \sigma_{\epsilon,i}^{2} \cdot Y^{\pi^0}_{i}
  	 /
         \sum_{i} Y^{\pi^0}_{i}$,
leads to 1\% uncertainty in the $\pi^0$ multiplicity.
Combining these two numbers gives a 3\%
systematic uncertainty in the $\pi^0$ mean multiplicity measurement
due to the fit procedure.


\begin{figure}
  \centering
  \includegraphics[width=8.6cm]{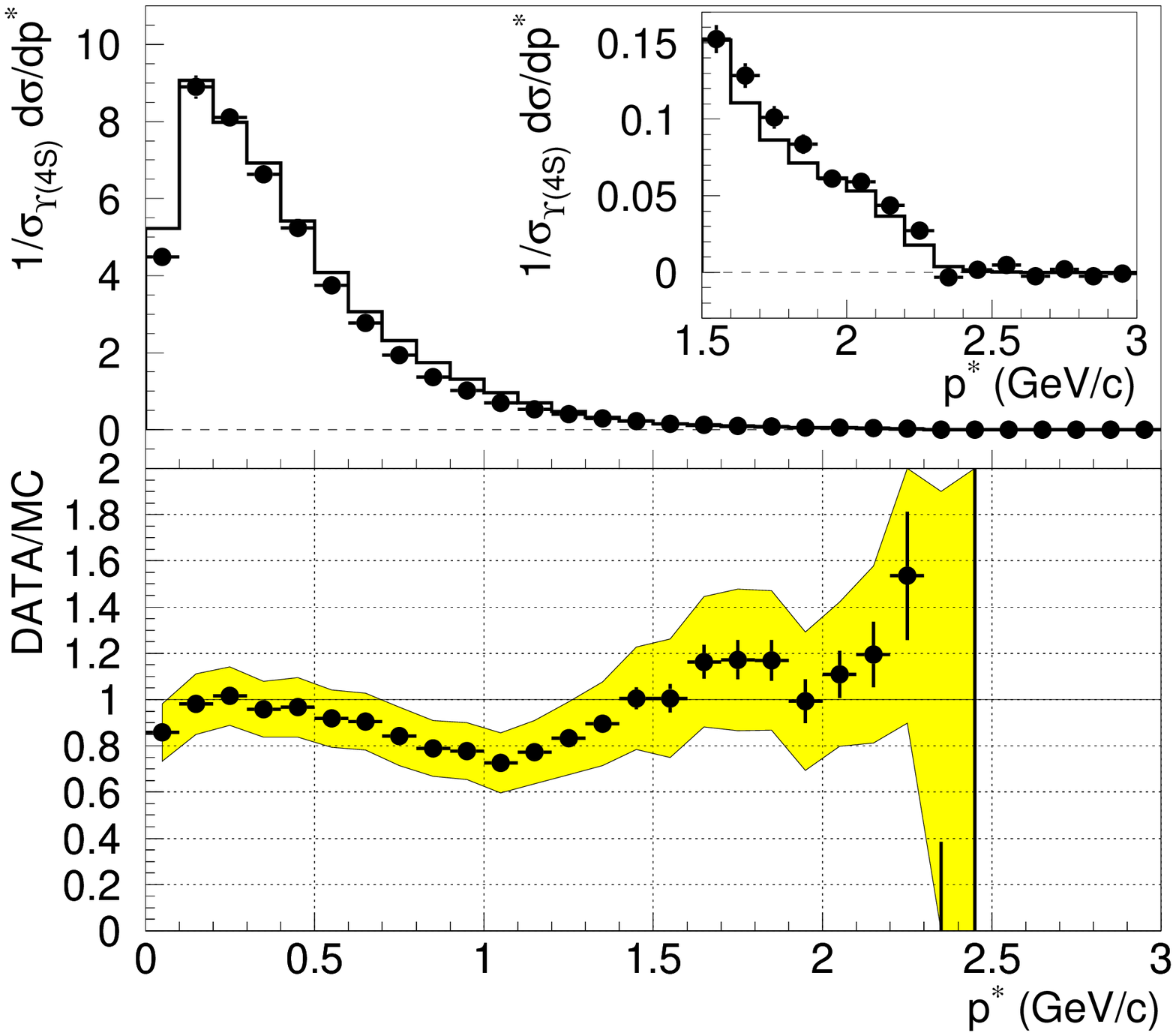}
  \caption{Measured inclusive $\pi^0$ momentum spectrum in $\Upsilon(4S)$ decays,
with the high momentum range shown in the inset.
	The histograms show the Monte Carlo prediction.
	The error bars are statistical, while the shaded band indicates
	one standard deviation systematic uncertainties.
	}
  \label{fig:data-qq}
\end{figure}

The measured $\pi^0$ momentum spectrum is compared to the $B\bar{B}$
Monte Carlo~\cite{ref-cleo-qq} prediction in Fig.~\ref{fig:data-qq}.
Systematic differences in the intermediate momentum region
may be caused by overestimation of $b \rightarrow c$ processes
in the Monte Carlo event generation: 100\% $b \rightarrow c$ is assumed
for generic $B\bar{B}$ decays in our Monte Carlo.
In the high momentum region above the $b \rightarrow c$ kinematic end point,
we searched for a $\pi^0$ excess from charmless $B$ decays.
We find
$N_{\rm excess} =  410 \pm 724$
events in the momentum interval 2.4--2.7$\,\GeVc$,
which corresponds to a partial branching fraction
${\cal B} (B \rightarrow \pi^0 X; \; p > 2.4\,\GeVc) = (2 \pm 3.5 \pm 6.7) \times 10^{-4}$,
or
less than $1.4\times10^{-3}$  
at 90\% confidence level.
However,
the statistical precision here is limited and the fluctuations
are dominated by the off-resonance data used in the subtraction.


In conclusion,
using 3.2 $\rm fb^{-1}$ of data
accumulated at the $\Upsilon(4S)$ resonance by the Belle detector,
we have measured the inclusive spectrum of neutral pions from
$\Upsilon(4S)$ decays.
By summing the measured momentum bins, the mean multiplicity
of neutral pions from $\Upsilon(4S)$ decays has been determined to be
$\langle n_{\pi^0} \rangle = 4.70 \pm 0.04 \pm 0.22$,
corresponding to an inclusive branching fraction\footnote{
Inclusive branching fractions in $B$ decays
have a definition in terms of multiplicity
and can be greater than 100\%~\cite{ref-pdg}.
}
of ${\cal B}(B \rightarrow \pi^0 X) = (235 \pm 2 \pm 11)$\%,
where the first error is statistical and the second is systematic.

\begin{acknowledgments}
%
%
We gratefully acknowledge the efforts of the KEKB group in providing
us with excellent luminosity and running conditions and the members
of the KEK computing research center
for their assistance with our computing and network systems.
We thank the staffs of KEK and collaborating institutions for their
contributions to this work,
and acknowledge support from the Ministry of Education, Science, Sports and
Culture of Japan and
the Japan Society for the Promotion of Science;
the Australian Research Council and the Australian Department of Industry,
Science and Resources;
the Department of Science and Technology of India;
the BK21 program of the Ministry of Education of Korea and
the Basic Science program
of the Korea Science and Engineering Foundation;
the Polish State Committee for Scientific Research 
under contract No.2P03B 17017; 
the Ministry of Science and Technology of Russian Federation;
the National Science Council and the Ministry of Education of Taiwan;
the Japan-Taiwan Cooperative Program of the Interchange Association;
and  the U.S. Department of Energy.
\end{acknowledgments}

%


\end{document}